\documentclass[%
preprint,
 showpacs,preprintnumbers,
amssymb,
 aip,
jcp,
 longbibliography
]{revtex4-1}

\usepackage{amsmath}
\usepackage{mathbbol}
\usepackage{bbold}
\usepackage{mathrsfs}
\usepackage{graphicx}
\usepackage{dcolumn}
\usepackage{bm}
\usepackage[usenames]{color}
\usepackage[normalem]{ulem}

\usepackage[per-mode=symbol]{siunitx}   
\usepackage[version=3]{mhchem}
 
\DeclareSIUnit\intensity{\watt\per\centi\meter\squared}
\DeclareSIUnit\fieldstrength{\volt\per\centi\meter}
\DeclareSIUnit\kfieldstrength{k\volt\per\centi\meter}
\DeclareSIUnit\energy{cm^{-1}}

\newcommand{\melement}[3]{\ensuremath{\left\langle #1 \left|#2\right|#3\right\rangle}}%
\newcommand{\ie}{i.\,e.}%
\newcommand{\singlets}{\ensuremath{{}^1\text{S}_0}}
\newcommand{\doubletshalf}{\ensuremath{{}^2\text{S}_{1/2}}}
\newcommand{\doublets}{\ensuremath{{}^2\text{S}}}
\newcommand{\doubletpodd}{\ensuremath{{}^2\text{P}^\text{o}}}

\newcommand{\ket}[1]{\left|#1\right\rangle}
\newcommand{\bra}[1]{\left\langle #1\right|}
\newcommand{\escalar}[2]{\left\langle #1|#2\right\rangle}

\newcommand{\aarphys}{\affiliation{Department of Physics and Astronomy, Aarhus University, 8000  Aarhus C, Denmark}}%
\newcommand{\uoft}{\affiliation{Chemical Physics Theory Group, Department of Chemistry,and Center for Quantum Information and Quantum Control, University of Toronto, Toronto, ON M5S 3H6, Canada}}

\begin{document}

\title{Effects of core space and excitation levels on ground-state correlation and photoionization dynamics of Be and Ne}

\author{Juan J.\ Omiste}\uoft
\author{Lars Bojer Madsen}\aarphys

\date{\today}
\begin{abstract} 
We explore the effects of correlation on the ground-state energies and on photoionization dynamics in atomic Be and Ne. We apply the time-dependent restricted-active-space self-consistent-field method for several excitation schemes and active orbital spaces with and without a dynamic core to address the effects systematically at different levels of approximation. For the ground-state many-electron wave functions, we compare the correlation energies with entropic measures of entanglement. A larger magnitude of the correlation energy does not always correspond to a larger value of the considered entanglement measures. To evaluate the impact of correlation in a process involving continua, we consider photoionization by attosecond pulses. The photoelecton spectra may be significantly affected by including a dynamical core.
\end{abstract}
\maketitle
\section{Introduction}
\label{sec:introduction}

Ultrafast processes in atomic and molecular systems have attracted the attention of many theoretical and experimental groups. For example, recent studies centered around laser-matter interaction have considered the measurement of time-delay in photoionization~\cite{Schultze2010,Klunder2011,Isinger2017}, dichroism~\cite{NgokoDjiokap2014}, chirality~\cite{Ordonez2018}, discrimination of chiral molecules using non-linearly polarized laser pulses~\cite{Ayuso2018}, spectroscopy and ultrafast molecular processes~\cite{Young2018, Calegari2014,Kraus2015science}.  The theoretical understanding of ultrafast electronic processes in atoms and molecules has been reached partly through the development of simple models such as, for example, the strong-field approximation~\cite{Keldysh1965,Faisal1973,Reiss1980}, and partly by the development of algorithms and theoretical methods, which aim at accurately describing many-electron quantum systems and accounting for the correlation among the electrons~\cite{Helgaker2000,Hochstuhl2014}. The effects of electron-electron correlation have been a central topic in atomic physics since the early works on the ground state energy in He~\cite{Hylleraas1928,Hylleraas1929}, and ground-state correlation is a central topic in quantum chemistry~\cite{Helgaker2000}.  In such time-independent settings, assessment of accuracy may benefit from the variational principle. The situation in the case of dynamics and in ultrafast processes is often less clear.  Studies of electron-electron correlation in a dynamical setting is therefore a central topic for current experimental and theoretical developments in ultrafast atomic, molecular and optical physics. For example recent works\cite{Gruson734,Kaldun738} have addressed the real-time correlated dynamics of doubly-excited Fano resonances~\cite{Fano} in He.   While a system like He can be dealt with by ab initio approaches directly solving the time-dependent Schr\"{o}dinger equation numerically, the situation is quite different for larger systems.  In such cases, and restricting the discussion to wave function based methodology, approaches involving orbitals have to be used with the mean-field time-dependent Hartree-Fock approach being the simplest example of such an orbital-based approach. Depending on the observable of interest, it is often only computationally possible to account for  a relatively small amount of correlation. In particular because continua requiring large simulation volumes are often involved. Indeed, progress has been impeded by the computational challenges imposed by the exponential scaling of the size of the problem with the number of electrons, and also by the lack of efficient ways to gradually improve the level of approximation. Therefore to address part of these questions,  we present in this work a systematic study of correlations in atomic Be and Ne for the ground states as well as for attosecond photoionization dynamics involving continua. 

In general,  if $M$ parameters are needed to describe each particle, the computational complexity of the problem scales as $M^N$ with $N$ the number of electrons - and this for each time-step. To reduce the base in this scaling. self-consistent field (SCF) methods have been considered. In these methods the number of orbitals can be reduced by introducing time-dependent orbitals, which therefore implies a smaller number of configurations. The most commonly used SCF methods which include correlation are multiconfigurational time-dependent Hartree-Fock (MCTDHF)~\cite{Beck2000a,Caillat2005,Nest2005,Remacle2007,Kato2008,Kato2008a,Kato2009,Hochstuhl2010,Haxton2011,Haxton2012,Haxton2015,Hochstuhl2014,Alon2007}, time-dependent complete-active-space SCF (TD-CASSCF)~\cite{Sato2013, Miyagi2014b,Sato2015,Sato2016a,Orimo2018} and time-dependent restricted-active-space SCF (TD-RASSCF)~\cite{Miyagi2013,Miyagi2014,Miyagi2014b,Miyagi2017,Omiste2017_be,Omiste2018_neon,Madsen2018} approaches. Specifically, TD-RASSCF not only reduces the number of orbitals, but also the number of configurations can be reduced by appropriately choosing the restricted-active-space (RAS). Note that these methods have been successfully developed for bosons also~\cite{Alon2007,Leveque2016a,Leveque2018}.  We will use the TD-RASSCF in this work. The effect of space partition and excitation scheme on the ground state correlation and the dynamics in the continuum can now be systematically explored. In addition, we find it of interest to explore to what extend trends in the correlation energy are reflected in entanglement measures. In quantum mechanics, the entanglement among particles plays a major role not only in the ground-state of the system, but also in the dynamics induced by, for instance, an external laser pulse. While the correlation energy is a very good measure for the ground state correlation, there is no corresponding measure for the case of dynamics. Accordingly, this part of our study is motivated by the expectation that  if for the ground states, entanglement measures reflect the correlation energy, i.e., if they increase when the magnitude of the correlation energy increases, then the former measures show the desired trend, and they could be expected to work also for dynamics and continua where they can also be evaluated. Hence in this way convergence of an entanglement measure with basis, orbitals and excitation scheme may be used as a means to identify the most important physics and to access the quality of a given calculation involving continua.

In many-electron systems, the degree of entanglement can be quantified by means of the trace of functions of the one-body density, $\rho$, through measures such as the linear or von Neumann entropy~\cite{Plastino2009}.  The practical evaluation of these quantities depends on the method used to describe the many-body system, and can become unaffordable for methods with a large number of orbitals, such as the time-dependent configuration interaction with singles (TD-CIS)~\cite{Karamatskou2014,Karamatskou2017,Toffoli2016,Sato2018}, the time-dependent general-active-space configuration interaction (TD-GASCI)~\cite{Bauch2014,Chattopadhyay2015,Yue2017} and time-dependent restricted-active-space configuration interaction~\cite{Hochstuhl2012,Hochstuhl2013} methods. In these methodologies the orbitals used to build the configurations do not adapt during the propagation and, therefore, a large number of orbitals and configurations is required to describe a given many-body quantum state. This demand, in turn, makes the calculation of the entropies and therefore the estimation of the degree of entanglement inefficient. On the other hand, because of the reduction in orbitals and configurations obtained by the use of time-dependent SCF orbitals, these methods ensure that the entropy and hence the entanglement of the system can be efficiently evaluated and that the main properties of the system may at the same time be accurately described due to the adaptability of the orbitals. 

The paper is organized as follows. In Sec.~\ref{sec:theory}, we describe the Hamiltonian and the TD-RASSCF method as well as the entanglement measures to be used. In Sec.~\ref{sec:results}, we present the results for Be and Ne, regarding the ground-state energy (Sec.~\ref{sec:gs_energy}), their entanglement (Sec.~\ref{sec:gs_entanglement}) and the photoelectron spectrum (PES) (Sec.~\ref{sec:gs_photoelectron}). Section~\ref{sec:conclusions} summarizes the main findings and concludes. Atomic units ($\hbar=m_e=e=a_0=1$) are used throughout unless indicated otherwise.

\section{Theory}
\label{sec:theory}

In this section we briefly describe the TD-RASSCF method as well as the correlation energy and the entropy-based entanglement measures used to quantify the correlation in Be and Ne.

\subsection{TD-RASSCF method}
\label{sec:TD-RASSCF method}
\begin{figure*}
\centering
    \includegraphics[width=.96\linewidth]{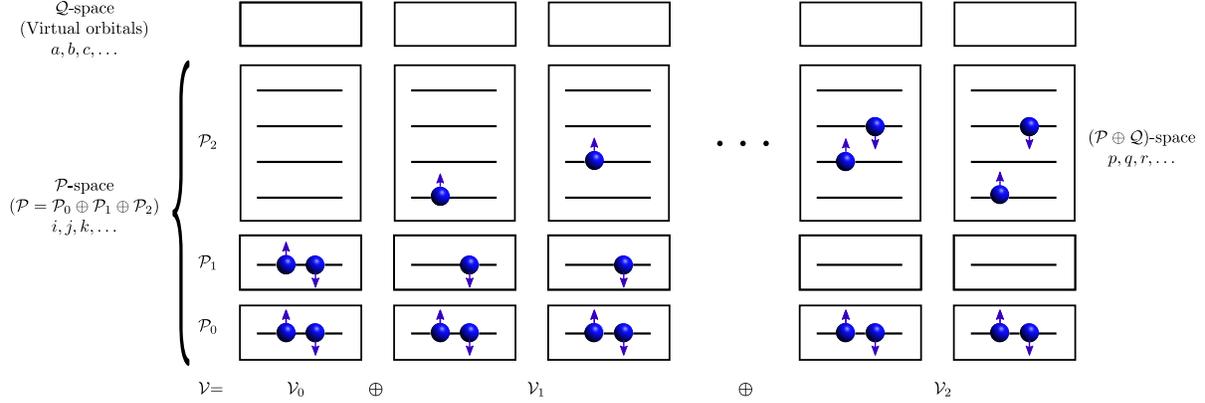}
    \caption{\label{fig:fig1} Illustration of the Fock-space for the TD-RASSCF method for Be. In the figure, the number of spatial orbitals in the three subspaces $ \mathcal{P}_0, \mathcal{P}_1$ and $\mathcal{P}_2$ are $M_0=1,M_1=1$ and $M_2=4$, respectively. Each subspace~$\mathcal{V}_N$ contains all the possible $N$ excitations from $\mathcal{P}_1$ to $\mathcal{P}_2$. The TD-RASSCF-S method includes the subspaces $\mathcal{V}_0\oplus\mathcal{V}_1$, TD-RASSCF-D includes $\mathcal{V}_0\oplus\mathcal{V}_2$ and TD-RASSCF-SD method $\mathcal{V}_0\oplus\mathcal{V}_1\oplus\mathcal{V}_2$. Note that TD-RASSCF-SD for $M=6$ and a single spatial orbital in the $\mathcal{P}_0$ core space is equivalent to TD-CASSCF, since all the possible configurations given by the orbitals in $\mathcal{P}_1\oplus\mathcal{P}_2$ are included.}
  \end{figure*}

The TD-RASSCF method is used to propagate the  many-electron wave function. The method was described in detail elsewhere~\cite{Miyagi2013, Miyagi2014,Miyagi2014b,Omiste2017_be, Omiste2018_neon,Madsen2018}, so the description here will be brief.  The TD-RASSCF methodology is a generalization of MCTDHF~\cite{Beck2000a,Koch2006} obtained by dividing the active orbital space, $\mathcal{P}$, into three subspaces, and by imposing restrictions on the excitations between them~\cite{Miyagi2013,Miyagi2014b}, \ie, the Ansatz of the many-body wave function reads
\begin{equation}
  \label{eq:wf_ansatz}
  \ket{\Psi(t)}=\sum_{\mathbf{I}\in\mathcal{V}} C_\mathbf{I}(t) \ket{\Phi_\mathbf{I}(t)},
\end{equation}
where the sum runs over the set of configurations $\mathcal{V}$, and $C_\mathbf{I}(t)$ and $\ket{\Phi_\mathbf{I}(t)}$ are the amplitudes and Slater determinants of the configuration $\mathbf{I}$, which are direct products of spin-up and spin-down strings, i.e., $\mathbf{I}=\mathbf{I_\uparrow}\otimes\mathbf{I_\downarrow}$, each of them including the indices of the spatial orbitals~\cite{Olsen1988,Klene2003}. Each Slater determinant is built from time-dependent spatial orbitals~$\{\phi_j(t)\}_{j=1}^M$. In the case of MCTDHF, $\mathcal{V}\equiv \mathcal{V}_\textup{FCI}$, that is, the full configuration space~\cite{Koch2006}. On the other hand, in the case of TD-RASSCF~\cite{Miyagi2014b,Miyagi2013}, the configurations run in the restricted active space, $\mathcal{V}\equiv \mathcal{V}_\textup{RAS}$, which is defined by the restrictions on the excitations in the active space. The active orbital space $\mathcal{P}$ is divided into 3 subspaces:~$\mathcal{P}_0,\,\mathcal{P}_1$ and $\mathcal{P}_2$, as illustrated in Fig.~\ref{fig:fig1}. $\mathcal{P}_0$ constitutes the core, and its orbitals are fully occupied at all times. Still these orbitals are time-dependent and describe, e.g., polarization of the core. All the combinations of the orbitals in $\mathcal{P}_1$ are allowed and the number of occupied orbitals in $\mathcal{P}_2$ correspond to the permitted excitations from $\mathcal{P}_1$. 
In this work, we apply TD-RASSCF including single (-S), double (-D) and single and double excitations (-SD) from the active space partition $\mathcal{P}_1$ to $\mathcal{P}_2$, with and without orbitals in the $\mathcal{P}_0$ core space [see Fig.~\ref{fig:fig1}].

 The TD-RASSCF theory is conveniently formulated in second quantization. We work in the spin-restricted framework, which implies that a given configuration $\ket{\Phi_\mathbf{I}(t)}$, describing $N_e$ electrons, is constructed by $N_e/2$ spatial orbitals. The Hamiltonian of a many-electron atom in the presence of a laser field reads 
\begin{equation}
  \label{eq:hamil_second}
H= \sum_{p,q} h^p_q(t)E^q_p +\frac{1}{2}\sum_{pqrs}v_{qs}^{pr}(t)E^{qs}_{pr},
\end{equation}
where the spin-free excitation operators $E^q_p$ and $E^{qs}_{pr}$ are defined as
\begin{equation}
  \label{eq:epq}
E^q_p=\sum\limits_{\sigma=\uparrow,\downarrow}b^\dagger_{p,\sigma} b_{q,\sigma}
\end{equation}
and
\begin{equation}
  \label{eq:epqrs}
E^{qs}_{pr}=\sum\limits_{\sigma=\uparrow,\downarrow}\sum\limits_{\gamma=\uparrow,\downarrow}b^\dagger_{p,\sigma}b^\dagger_{r,\gamma} b_{s,\sigma} b_{q,\gamma}
\end{equation}
with $b^\dagger_{p,\sigma}$ and $b_{p,\sigma}$ the creation and annihilation operators of a single spin-orbital $\ket{\phi_{p}(t)}\otimes\ket{\sigma}$. In Eq.~\eqref{eq:hamil_second}, the one-body, $h_q^p(t)$, and two-body, $v_{qs}^{pr}(t)$ matrix elements, are given by
\begin{eqnarray}
h_q^{p}(t)&=& \int\mathrm{d}\vec{r}\phi_p^\star(\vec{r},t)h(\vec{r},t)\phi_q(\vec{r},t),\\
\label{eq:vqspr}
v_{qs}^{pr}(t)&=&  \int\int\mathrm{d}\vec{r}\mathrm{d}\vec{r}'\cfrac{\phi_p^\star(\vec{r},t)\phi_r^\star(\vec{r}',t)\phi_q(\vec{r},t)\phi_s(\vec{r}',t)}{|\vec{r}-\vec{r}'|},
\end{eqnarray}
where 
\begin{equation}
\label{eq:h_1st_quantization}
  h(\vec{r},t)= \cfrac{p^2}{2}-\cfrac{Z}{r}+V_L(\vec{r},t).
\end{equation}
Here $V_L(\vec{r},t)$ is the interaction with the laser field taken to be in the dipole approximation
\begin{equation}
  \label{eq:lv_gauge}
V_L(t,\vec{r})=\left\{
\begin{array}{ll}
  \vec{E}(t)\cdot \vec{r},&~\text{Length gauge (LG)}\\
-i \vec{A}(t)\cdot\vec{\nabla},&~\text{Velocity gauge (VG).}
\end{array}
\right.
\end{equation}
In Eq.~\eqref{eq:lv_gauge} $\vec{E}(t)$ is the electric field of the laser and is obtained as the time derivative of the vector potential $\vec{E}(t)=-\partial_t \vec{A}(t)$~\cite{Han2010}. One of the main characteristics of the TD-RASSCF method is its invariance under gauge transformations~\cite{Miyagi2014b}. In this work all calculations have been checked to give identical results in the two gauges.

 The equations of motion (EOM) determining the time evolution of the amplitudes $C_\mathbf{I}(t)$ and the orbitals $\ket{\phi_i(t)}$ may be written as~\cite{Miyagi2013}
\begin{widetext}
  \begin{align}
    \label{eq:ci_dot}
    &i\dot C_\mathbf{I}(t)=\sum_{ij}\left[h^i_j(t)-i\eta_j^i(t)\right]\langle\Phi_\mathbf{I}(t)|E_i^j|\Psi(t)\rangle+\cfrac{1}{2}\sum_{ijkl}v_{jl}^{ik}(t)\langle\Phi_\mathbf{I}(t)|E_{ik}^{jl}|\Psi(t)\rangle,\\
    \label{eq:q_space}
    &i\sum_j Q(t) |{\dot\phi_j(t)}\rangle\rho_i^j(t)=\sum_jQ(t)\left[h(t)|{\phi_j(t)}\rangle\rho_i^j(t)+\sum_{jkl}W_l^k(t)|{\phi_j(t)}\rangle\rho_{ik}^{jl}(t)\right],\\
    \label{eq:p_space}
    &\sum_{k''l'}\left[h_{l'}^{k''}(t)-i\eta_{l'}^{k''}(t)\right]A_{k''i'}^{l'j''}(t)+\sum_{klm}\left[v_{kl}^{j''m}(t)\rho_{i'm}^{kl}(t)-v_{i'm}^{kl}(t)\rho_{kl}^{j''m}(t)\right]=i\dot\rho_{i'}^{j''}(t), 
  \end{align}
\end{widetext}
where the orbitals with single and double prime belong to different subspaces, and where the following quantities were introduced
\begin{widetext}
  \begin{eqnarray}
    \eta_j^{i}(t)&=& \langle \phi_i(t)|\dot{\phi}_j(t)\rangle,\, Q(t)=\mathbb{1}-P(t)=\mathbb{1}-\sum_{j=1}^M \ket{\phi_j(t)}\bra{\phi_j(t)}\\
\label{eq:e_rho}
    E_p^q&=&c_p^\dagger c_q,\quad E_{pr}^{qs}=c_p^\dagger c_r^\dagger c_q c_s,\quad \rho_i^j(t)=\langle\Psi(t)|c_i^\dagger c_j|\Psi(t)\rangle,\\
    \label{eq:e_e_rho}
    \rho_{ik}^{jl}(t).&=&\langle\Psi(t)|c_i^\dagger c_k^\dagger c_l c_j|\Psi(t)\rangle,\quad A_{ki}^{lj}(t)= \langle\Psi(t)|[c_i^\dagger c_j, c_k^\dagger c_l]|\Psi(t)\rangle.
    \label{eq:rho_a}
  \end{eqnarray}
\end{widetext}
In a schematic manner, the time dependence of the amplitudes, $C_\mathbf{I}(t)$, and time-dependent spatial orbitals, $\ket{\phi_j(t)}$ fulfil the non-linear equations
\begin{eqnarray}
  \label{eq:ci_dot_formal}
  \dot{C}_\mathbf{I}(t)&=&f_\mathbf{I}(\{C_\mathbf{J}(t)\}_\mathbf{J},\{\phi_k(t)\}_k)\\
\label{eq:phi_dot_formal}
  \ket{\dot\phi_k(t)}&=&\sum\limits_l\eta_{k}^l\ket{\phi_l(t)}+Q(t)\ket{\dot\phi_k(t)},
\end{eqnarray}
where the functions, $f_\mathbf{I}$, of Eq.~\eqref{eq:ci_dot_formal} are given explicity in Eq.~\eqref{eq:ci_dot}. To solve these equations we first compute $Q(t)\ket{\dot\phi_k(t)}$ and $\eta_{k}^l$  using Eqs.~\eqref{eq:q_space} and~\eqref{eq:p_space}, respectively. Then, by plugging these quantities into Eq.~\eqref{eq:phi_dot_formal} we compute $\ket{\dot\phi_k(t)}$. The amplitudes ${C}_\mathbf{I}(t)$ are obtained by solving Eq.~\eqref{eq:ci_dot}. Note that Eqs.~\eqref{eq:ci_dot_formal} and~\eqref{eq:phi_dot_formal} can be computed independently at each time step. For the time-propagation we use a Runge-Kutta method, but other methods, such as the Adams-Bashforth-Moulton method~\cite{Leveque2016a} can also be applied.
The main obstacle of this computation is the solution of Eq.~\eqref{eq:p_space} for $\eta_k^l$, since it implies the calculation of the time derivative of the one-body density operator. In the case of MCTDHF, this equation becomes an identity~\cite{Miyagi2013}, and we only need to solve Eqs.~\eqref{eq:ci_dot} and~\eqref{eq:q_space}. Furthermore, if we only allow double (or even) excitations from $\mathcal{P}_1$ to $\mathcal{P}_2$, Eq.~\eqref{eq:p_space} simplifies to the solvable form~\cite{Miyagi2013, Miyagi2014b, Leveque2016a, Omiste2017_be, Omiste2018_neon}
\begin{widetext}
  \begin{equation}
    \label{eq:p_space_doubles}
    \sum_{k''l'}\left[h_{l'}^{k''}(t)-i\eta_{l'}^{k''}(t)\right]A_{k''i'}^{l'j''}(t)+\sum_{klm}\left[v_{kl}^{j''m}(t)\rho_{i'm}^{kl}(t)-v_{i'm}^{kl}(t)\rho_{kl}^{j''m}(t)\right]=0.
  \end{equation}
\end{widetext}
We also note that $\dot\rho_{i'}^{j''}(t)=0$ if $i'\in\mathcal{P}_0$ and $j''\in\mathcal{P}_1\oplus\mathcal{P}_2$. In any other case, we use that Eq.~\eqref{eq:p_space} is equivalent to\cite{Miyagi2014b}
\begin{equation}
  \label{eq:p_space_tensorial}
  \sum_{k''l'}\left(i\eta_{l'}^{k''}(t)-h_{l'}^{k''}(t)\right)\xi_{k''i'}^{l'j''}=\frac{1}{2}\sum_{klmn}v_{ln}^{km}(t)\xi_{kmi'}^{lnj''}(t),
\end{equation}
where 
\begin{align}
  \xi_{k''i'}^{l'j''}(t)&= \melement{\Psi_{i'}^{j''}(t)}{(1-\Pi(t)E_{k''}^{l'})}{\Psi(t)}\\
  \xi_{kmi'}^{lnj''}(t)&= \melement{\Psi_{i'}^{j''}(t)}{(1-\Pi(t)E_{km}^{ln})}{\Psi(t)}
\end{align}
and $\Pi(t)=\sum\limits_{\mathbf{I}\in\mathcal{V}}\ket{\Phi_\mathbf{I}(t)}\bra{\Phi_\mathbf{I}(t)}$. Note that Eq.~\eqref{eq:p_space_tensorial} is solvable for the unknown $\eta$'s using the amplitudes and time-dependent orbitals at time $t$. 

The great advantage of the TD-RASSCF approach is that by choosing an appropriate RAS scheme for the physical system and process of interest it is possible to obtain both efficiency and accuracy. On the other hand, we also have to consider that at every time-step the orbitals and related operators must be updated. This computation may become unaffordable when considering processes involving continua of a many-electron system, mainly due to the high number of operations needed to obtain the two-body matrix elements $v_{qs}^{pr}(t)$ of Eq.~\eqref{eq:vqspr}. To speed up the evaluation of these matrix elements, we use the coupled basis method~\cite{Omiste2017_be,Omiste2018_neon}. This method uses that  it is more efficient to couple the angular momenta of the orbitals before computing the two-body matrix elements than performing the calculation directly and benefits from the fact that the electron repulsion operator commutes with the angular momentum of two electrons, $\left[\dfrac{1}{|\vec{r}_1-\vec{r}_2|},(\vec{\ell}_1+\vec{\ell}_2)^2\right]=\left[\dfrac{1}{|\vec{r}_1-\vec{r}_2|},\ell_{1,z}+\ell_{2,z}\right]=0$. Finally, our implementation also uses properties of the finite element discrete variable representation (FE-DVR) grid for the radial part~\cite{Omiste2017_be}. 

\subsection{Correlation and entanglement}
\label{sec:correlation_and_entanglement}

In this section we describe the quantities that are used to compute the degree of correlation at the different levels of approximation of the RAS schemes. The standard measure for correlation in a many-body wave function is given in terms of the \emph{correlation energy}, $E_\text{corr}$, defined as the difference between the total energy and the Hartree-Fock energy~\cite{Lowdin1955,PerOlov1995}. In Sec.~III, we study the effects on the correlation energy of the partition of the orbital space and the excitation level (singles, doubles, single-doubles). We also find it interesting to investigate to which extend a given trend in the correlation energy is reflected in the degree of quantum mechanical entanglement. We would expect a higher degree of entanglement, the larger the magnitude of the correlation energy. Entanglement is associated with the non-separability of a many-body wave function. Two particles are entangled if their associated wave function cannot be written as a product of single-particle wave functions $\ket{\Psi(r_1,r_2)}=\ket{\xi(r_1)}\otimes\ket{\chi(r_2)}$. According to this criterium, the fermionic nature of electrons implies that the total electronic wave function for atoms and molecules is always entangled. Therefore, the notion of entanglement is meaningfully extended to fermions by defining a wave function as being entangled if it can not be written as a single Slater determinant~\cite{Schliemann2001}. The quantification of entanglement and correlation is a difficult task~\cite{Eckert2002,Amico2008,Plastino2009}. For two particles the degree of entanglement is quantified by the purity $P(\rho)=\text{Tr}\left\{\rho^2\right\}$ and the von Neumann entropy, $S(\rho)=-\text{Tr}\left\{\rho\ln\rho\right\}$, with $\rho$ the density matrix of the considered two-particle system. Physically, the purity $P(\rho)$ in the density matrix formalism quantifies how pure a state of two particles is, being maximum for $\text{Tr}\left\{\rho^2\right\}=1$ and minimum for a maximally mixed state, $\text{Tr}\left\{\rho^2\right\}=1/d$ with $d$ the dimension of the Hilbert space~\cite{Nielsen2010}. On the other hand, the von Neumann entropy is a generalization of the Shannon entropy to quantum mechanics, being $0$ for a pure state and $\ln d$ for a maximally mixed state. Classically, the Shannon entropy quantifies the uncertainty we have of a given variable before we measure it or, in other words, the information we obtain after measuring the variable~\cite{Nielsen2010}. To quantify the amount of entanglement in an $N$-body wave function we use the two measures~\cite{Plastino2009,Majtey2016}
\begin{subequations}
  \begin{equation}
    \label{eq:el}
    \mathscr{E}_L(\rho) =2N-\text{Tr}\left\{\rho^2\right\},
  \end{equation}
  \begin{equation}
    \label{eq:evn}
    \mathscr{E}_{VN}(\rho) =-\text{Tr}\left\{\rho\ln\rho\right\}+N\ln 2.
  \end{equation}
\end{subequations}
Here the elements of the density matrix $\rho$ are given in Eq.~\eqref{eq:e_rho}.  The subscripts reflect that the measures are derived from the linear and the von Neumann entropies, respectively. The measures $\mathscr{E}_L(\rho)$ and $\mathscr{E}_{VN}(\rho)$ are positive and Eqs.~\eqref{eq:el} and ~\eqref{eq:evn} are defined such that if the wave function can be written as a single Slater determinant, the entanglement is zero, $\mathscr{E}_L(\rho)=\mathscr{E}_{VN}(\rho)=0$~[\onlinecite{Plastino2009}]. Note that our definition differs from the definition in Ref.~[\onlinecite{Plastino2009}] since there the trace of the one-body density is normalized to unity. In our case, however, $\text{Tr}(\rho)=N$ and we use spin-free operators to compute $\rho$ in Eq.~\eqref{eq:epq} instead of spin-specific creation and anihilation operators. The  maximum values of the measures we use are
\begin{subequations}
  \begin{equation}
    \label{eq:el_max}
    \mathscr{E}_L(\rho)_\text{max} =2N\left(1-\frac{N}{2M}\right),\\
  \end{equation}
  \begin{equation}
    \label{eq:evn_max}
    \mathscr{E}_{VN}(\rho)_\text{max} =N\ln\left(\frac{2M}{N}\right).
  \end{equation}
\end{subequations}
where $M_\text{}$ is the number of spatial orbitals, with $M_\text{}\ge\frac{N}{2}$. Let us note that the entanglement measures~\eqref{eq:el} and~\eqref{eq:evn} are derived from separability criteria and a more elaborated measure would be required to exactly describe the entanglement~\cite{Majtey2016}. The efficient computation of the entanglement measures $\mathscr{E}_L(\rho)$ and $\mathscr{E}_{VN}(\rho)$ constitute a great advantage in contrast to more sophisticated proposed measures for $N$-particle systems~\cite{Amico2008,Majtey2016}, and the relatively small size of $\rho$ in the case of TD-RASSCF makes the evaluation easier than in the case of non-SCF methods, where typically many more orbitals are needed.

\section{Results}
\label{sec:results}

In this section we compute the absolute ground-state energy of Be and Ne using the TD-RASSCF method including single (-S), double (-D) and single and double (-SD) excitations with and without a core of$~\mathcal{P}_0$ orbitals~\cite{Miyagi2013,Miyagi2014b} [Fig. 1]. The ground-state is computed performing imaginary time propagation (ITP) on an initial guess function. This guess function can be chosen to be fully random, but this approach was shown to be most suitable for simple systems such as He~\cite{Hochstuhl2010}. For more complex systems a designed guess wave function where the initial orbitals and configurations are chosen to reflect the symmetry of the ground-state is more suitable in terms of convergence~\cite{Omiste2017_be}. In this work, we take the initial orbitals to be hydrogenic with nuclear charge $Z$ and ordered as $1s,\, 2s,\, 2p,\, 3s,\, 3p$ and $3d$. This selection is adequate since the magnetic quantum number of each orbital is preserved in the MCTDHF, TD-CASSCF and TD-RASSCF methods~\cite{Sato2016a, Omiste2018_neon}. See Appendix~\ref{sec:conservation_m} for further details. To choose the initial values for the amplitudes $C_\mathbf{I}(t)$, we consider the symmetry of the ground-state, which is $^1\text{S}^e$ for both Be and Ne. Accordingly we set the total magnetic quantum number $M_L=0$ and the total parity to be even in the guess function. To do so, the amplitude $C_\mathbf{I}(t)\ne 0$ when $M_L=0$ of the considered configuration $\mathbf{I}$~\cite{Omiste2018_neon} and if its parity is even, that is to say, $\sum_{i}\ell_i$ is even, with $\ell_i$ the angular momentum quantum number of the $i$th orbital. In the following calculations we use the same radial grid as in Refs.~\onlinecite{Omiste2017_be,Omiste2018_neon}. Specifically, to compute the ground-state of Be we use a radial box with $r\in[0,31]$, with 8 equidistant elements for $0\le r\le 8$ and 5 equidistant elements for $8\le r\le 28$. For the real time propagation we add 35 equidistant elements up to $r_\text{max}=200$. In the case of Ne, for the ground-state calculation $r\in[0,31]$, we use 12 equidistant elements for $0\le r\le 6$ and 10 equidistant elements for $6\le r\le 31$. We complete the radial box with 68 equidistant elements until the radial distance $r_\text{max}=201$. The time-steps used range from $10^{-4}$ to $10^{-3}$.

\subsection{Energy}
\label{sec:gs_energy}

In this section we compute the ground-state energies of Be and Ne by propagating the TD-RASSCF equations in imaginary time~\cite{Miyagi2013, Miyagi2014, Miyagi2014b,Omiste2017_be,Omiste2018_neon}. In Tables~\ref{tab:gs_energy_be_core} and~\ref{tab:gs_energy_be_nocore} we show the ground-state energies of Be for several RAS schemes, together with the number of configurations. Note that for Be we do not consider single excitations, because the associated one-body density matrix will be singular by construction for the initial guess functions used, leading to inconsistencies in Eq.~\eqref{eq:q_space} since the inverse does not exist. If $\rho$ is singular then at least one of its eigenvalues is zero, in other words, at least one of the orbitals is not used. As we described in the previous section, the orbitals in the guess function in the case of Be are $1s$, $2s$ (which are occupied for the Hartree-Fock configuration), and $2p$, $3s$, $3p$ and $3d$. Any configuration including a single excitation from $1s$ or $2s$ to $2p$ is a $P$ term, which within the considered LS-coupling scheme does not couple to $\singlets$, that is, it cannot contribute to the ground-state. This means that these configurations are not included in the wave function, and, since the $2p$ orbitals only appear in this type of configurations, the $2p$ orbitals do not contribute in the TD-RASSCF-S method, leading to a singular density matrix. On the contrary, in the case of TD-RASSCF-D and TD-RASSCF-SD, all the orbitals contribute since they can couple to the \singlets. Therefore, the density matrix $\rho$ is not singular in general. Let us remark that the orbitals develop into nonhydrogenic orbital during propagations but we will continue labelling them as $1s$, $2s$, $2p$, $3s$, $3p$ and $3d$ for convenience.

First, we consider one spatial orbital, $1s$, in the core $\mathcal{P}_0$ which is present in all the configurations. The ground-state energies are shown in Table~\ref{tab:gs_energy_be_core}. We denote by $M$ the number of spatial orbitals used to construct the configurations. The number $M$ fulfils $M=M_0+M_1+M_2$, where $M_0$, $M_1$ and $M_2$ are the number of spatial orbitals in $\mathcal{P}_0$, $\mathcal{P}_1$ and $\mathcal{P}_2$ subspaces. The notation $(M_0,M_1,M_2)$ specify the orbitals together with the excitation levels. We observe that for $M=5$, the ground-state energies are the same up to the fifth decimal digit for TD-RASSCF-D and TD-CASSCF, and they do not differ in the seventh first decimals for $M=6,9$ and $14$. In the TD-RASSCF notation, the TD-CASSCF refers to $M_0\ne 0$, $M_2=0$ and $M_1\ge N_e/2$. This slight improvement of the energy is due to a small number of configurations which are included in the TD-CASSCF but not considered in TD-RASSCF-D,~\ie, single excitations. For instance, TD-CASSCF with $M_\text{}=14$ requires only 169 configurations, $14$ more than TD-RASSCF-D. On the other hand, the ground-state energy converges very fast with the number of orbitals to the exact ground-state energy~\cite{Morrison1987}: the TDHF energy is $0.3\%$ higher than TD-CASSCF $(1,4,0)$, however, from $(1,4,0)$ to $(1,13,0)$ the ground-state energy diminishes only by $0.015\%$. The lack of configurations involving SD excitations from the $1s$ orbital is responsible for a lowest bound, estimated around $-14.6192$ [\onlinecite{Morrison1987}], which is quite different from the acurate ground-state energy $-14.6674$ given in Refs.~[\onlinecite{Lindroth1992},\onlinecite{Fischer1993}]. Physically these results can be understood by noting that the small nuclear charge is not able to strongly bind the electrons close the nucleus, therefore, the description with a dynamical $\mathcal{P}_0$ core subspace is not adequate to accurately describe the ground-state. 

In the case of the ground-state energies for $M_0=0$, shown in Table~\ref{tab:gs_energy_be_nocore}, excitations from $1s$ are allowed, leading to an improvement of $0.087\%$ of the ground-state energy for MCTDHF when the number of orbitals is increased from 5 to 6. We observe that the TD-RASSCF including -D and -SD differ only slightly. This is because the number of configurations is similar for both schemes. Note that eventhough the number of configurations is similar for -S and -SD in the considered case, the number of operations required in the -SD case is much larger than for -D~\cite{Miyagi2014b}.

As described at the beginning of Sec.~\ref{sec:results}, the initial guess functions is chosen such that all configurations and orbitals have a well defined parity and magnetic quantum number, which restricts the family of solutions and ensures that the ground-state wave functions have well defined parity and magnetic quantum number~\cite{Omiste2018_neon}. This implies that the ground-state energy obtained may be an upper bound of the absolute ground-state energy using this RAS scheme~\cite{Bardos2009}. For instance, the ground-state energy for MCTDHF with 5 orbitals is $-14.6171$, larger than the value $-14.6192$ in Ref.~[\onlinecite{Omiste2017_be}] which was taken as the most bound ground-state solution without taking into account any symmetry considerations. In the present work, we consider ground-state wave functions belonging to the same family of solutions,~\ie, functions with even parity, $M_L=0$ for each configuration and orbitals with well defined magnetic quantum number. As explained in Refs.~[\onlinecite{Omiste2017_be},\onlinecite{Bardos2009}], MCTDHF may have more than one local minimum energy solution. By setting the same symmetries on configurations and orbitals we ensure a physical meaning of the wavefunction (orbitals and conserved magnetic quantum number) and stability during the time propagation.

\begin{table*}\centering
\caption{\label{tab:gs_energy_be_core} Ground-state energies in atomic units of Be for several RAS schemes with one spatial orbital in the core $\mathcal{P}_0$ subspace,~\ie, $M_0=1$. $M$ denotes the number of spatial orbitals. The parenthesis in the left column specify $(M_0,M_1,M_2)$ with $M=M_0+M_1+M_2$. The integers in parentheses denote the number of configurations.}
\begin{ruledtabular}
  \begin{tabular}{cccccc}
     &  \multicolumn{5}{c}{$M_\text{}$} \\
     & 2  &5& 6  & 9&14 \\
    \hline
    $(1,\,1,\,M_\text{}-2)$ & - & -14.6171189  & -14.6186853  & -14.6187965  & -14.6192154 \\
    -D & & (10) & (17) & (50) & (145) \\
    \hline
    $(1,\,M_\text{}-1,\,0)$ & -14.5732980  & -14.6171191  & -14.6186853  & -14.6187965  & -14.6192154 \\
    TD-CASSCF & (1) & (16) & (25) &  (64) & (169)\\
\hline
Ref.~[\onlinecite{Morrison1987}] &   \multicolumn{5}{c}{-14.6192087}  \\
  \end{tabular}
\end{ruledtabular}
\end{table*}
\begin{table*}\centering
\caption{\label{tab:gs_energy_be_nocore} Ground-state energies in atomic units of Be for several RAS schemes without core $\mathcal{P}_0$ subspace,~\ie, $M_0=0$. $M$ denotes the number of spatial orbitals. The parenthesis in the left column specify $(M_0,M_1,M_2)$ with $M=M_0+M_1+M_2$. The integers in parentheses denote the number of configurations.}
\begin{ruledtabular}
  \begin{tabular}{cccccc}
     &  \multicolumn{5}{c}{$M_\text{}$} \\
     & 2  &5& 6  & 9&14 \\
    \hline
    $(0,\,2,\,M_\text{}-2)$ & - & -14.6171126  & -14.6297808  & -14.6508198   & -14.6555408\\
    -D & & (43) & (77) & (239) & (709)\\
    \hline
    $(0,\,2,\,M_\text{}-2)$ & - & -14.6171259  & -14.6311378  & -14.6508214  & -14.6555460\\
    -SD & & (53)  & (93)  & (267)  & (757)\\
    \hline
    MCTDHF & -14.5732980  &  -14.6171275 & 14.6311540 & -14.6541406 & -14.6590349 \\
    & (1) & (100) & (225) &  (1296) & (8281) \\
\hline
M=16 [\onlinecite{Hochstuhl2014}] &   \multicolumn{5}{c}{-14.659}  \\
\hline
Exact~\cite{Lindroth1992,Fischer1993} & \multicolumn{5}{c}{-14.6674}\\
  \end{tabular}
\end{ruledtabular}
\end{table*}
We now analyze the ground-state energy with and without a core for Ne. The results are collected in Tables~\ref{tab:gs_energy_ne_core} and~\ref{tab:gs_energy_ne_nocore}, respectively. The nuclear charge of Ne is much higher with $Z=10$ than for Be, which implies that the electrons in the $1s$ orbital are tightly bound. Therefore, we expect the ground-state energies to be similar in the case of the same RAS scheme for $\mathcal{P}_1$ and $\mathcal{P}_2$ but with $M_0=0$ and $1$, since the $1s$ electrons contribute essentially to all the relevant configurations. For instance, the schemes $(1,4,4)$ (Table~\ref{tab:gs_energy_ne_core}) and $(0,5,4)$ (Table~\ref{tab:gs_energy_ne_nocore}) differ by approximately $0.0018342$~a.u., that is, by $0.0014\%$. As opposed to the Be case for $M_0=1$, the ground-state energy decreases significantly as we increase the exitation level, due to the irrelevant role played by the excitations from the $1s$ orbital in the wave function. Thus, these two RAS schemes have approximately equivalent level of accuracy but the number of configurations for $(1,4,4)$ is $1.55$ times smaller that for $(0,5,4)$.

We further analyze the ground-state energy by setting $M_0=2$, that is, we do not allow excitation of the $1s$ and $2s$ orbitals. The $2s$ orbital is not as tightly bound as the $1s$ orbital, therefore, it has a noticeable effect on the ground-state energy, as we see in Table~\ref{tab:gs_energy_ne_core}. For instance, we find that for $M=9$ the ground-state energy for TD-CASSCF $(2,7,0)$ is approximately $0.037$~a.u. larger than for TD-CASSCF $(1,8,0)$, even more, it is just a bit more bound than for $(1,4,4)$-S, which only consists of 33 configurations. We find that it is very illustrative that the ground-state energy of RAS $(1,4,4)$-D is $-128.6773667$ with only 329 configurations, whereas TD-CASSCF $(2,7,0)$ results in $-128.6436521$ with almost 4 times more configurations.

\begin{table*}\centering
  \caption{\label{tab:gs_energy_ne_core} Ground-state energies in atomic units of Ne for several RAS schemes with one or two spatial orbitals in the core $\mathcal{P}_0$ subspace,~\ie, $M_0=1$ or 2. $M$ denotes the number of spatial orbitals. The parenthesis in the left column specify $(M_0,M_1,M_2)$ with $M=M_0+M_1+M_2$. The integers in parentheses denote the number of configurations. Underlined, $^\sharp$ and $^\flat$ denote that the energies are exactly identical~\cite{Miyagi2013,Miyagi2014,Miyagi2014b}. See text for further details.}
\begin{ruledtabular}
  \begin{tabular}{ccccc}
     &  \multicolumn{4}{c}{$M_\text{}$} \\
     & 5  &6& 9  & 14 \\
    \hline
    $(1,\,4,\,M_\text{}-5)$ & - & -128.5613201  & -128.6408287$^\flat$ & 128.6408287$^\flat$\\
    -S & & (9) & (33)  & (73) \\
    \hline
    $(1,\,4,\,M_\text{}-5)$ & - & \underline{-128.5613714}  & -128.6773667  & -128.7614146 \\
    -D & & (17) & (329)  & (1729) \\
    \hline
    $(1,\,4,\,M_\text{}-5)$ & - & \underline{-128.5613714}  & -128.6773826  & -128.7614302 \\
    -SD & & (25) & (361)  & (1801)\\
    \hline
     $(1,\,M_\text{}-1,\,0)$& -128.5481185 & \underline{-128.5613714}  & -128.6810539  &  - \\
    TD-CASSCF & (1) & (25) & (4900) & (511225) \\
    \hline
    $(2,\,3,\,M_\text{}-5)$ & - & {-128.5539044}$^\sharp$  & -128.6419084  & -128.7151655 \\
    -D & & (10) & (181)  & (946)\\
    \hline
     $(2,\,M_\text{}-2,\,0)$& -128.5481185 & {-128.5539044}$^\sharp$  & -128.6436521  &  -128.7181229 \\
    TD-CASSCF & (1)  & (16) & (1225) & (48400) \\
  \end{tabular}
\end{ruledtabular}
\end{table*}

Let us finally remark that the TD-RASSCF method satisfies several properties of the convergence of the energy with respect to the RAS scheme~\cite{Miyagi2014,Miyagi2014b} that we can extract from the results in Tables~\ref{tab:gs_energy_ne_core} and~\ref{tab:gs_energy_ne_nocore}. First of all, TD-RASSCF-SD and MCTDHF are the same method if $M_1=N_e/2$ and $M_2=1$, as it is shown for $(0,5,1)$ and MCTDHF with $M=6$. The same argument applies for  TD-RASSCF-SD and TD-CASSCF for $M_0+M_1=N_e/2$ and $M_2=1$. Moreover, the TD-RASSCF-D method is equivalent to MCTDHF for $M=N_e/2+1=6$, as it is the case for $(0,5,1)$, $(1,4,1)$ and $(2,3,1)$ and MCTDHF~\cite{Miyagi2014b,Omiste2018_neon}. Finally, the ground-state energies computed with TD-RASSCF-S for $(1,4,5)$ and $(1,4,9)$ are identical because the wave function is invariant with $M_2$ for $M_1\le M_2$, see Ref.~[\onlinecite{Miyagi2014}]. 
\begin{table*}\centering
\caption{\label{tab:gs_energy_ne_nocore} Ground-state energies in atomic units of Ne and number of configurations for several RAS schemes. $M$ denotes the number of spatial orbitals. The parenthesis in the left column specify $(M_0,M_1,M_2)$ with $M=M_0+M_1+M_2$. The integers in parentheses denote the number of configurations. Underlined energies are exactly identical~\cite{Miyagi2013,Miyagi2014,Miyagi2014b}. See text for further details.}
\begin{ruledtabular}
  \begin{tabular}{ccccc}
     &  \multicolumn{4}{c}{$M_\text{}$} \\
     & 5  &6& 9  & 14 \\
    \hline
    $(0,\,5,\,M_\text{}-5)$ & -  & -128.5613195  & -128.6408293  &  -128.6419576\\
    -S &  & (11) & (41)  & (91) \\
    \hline
    $(0,\,5,\,M_\text{}-5)$ & - & \underline{-128.5614288}  & -128.6792009  & -128.7649817 \\
    -D & & (26)  & (512)  & (2746) \\
    \hline
    $(0,\,5,\,M_\text{}-5)$ & - & \underline{-128.5614288}  & -128.6792085  & -128.7649890  \\
    -SD & & (36)  & (561) & (2836) \\
    \hline
    MCTDHF & -128.5481185  & \underline{-128.5614288} & -128.6828492  & -  \\
    & (1) & (36) & (15876)  & (4008004)\\
  \end{tabular}
\end{ruledtabular}
\end{table*}

\subsection{Entanglement}
\label{sec:gs_entanglement}

To see if there is a monotonic relationship between the correlation energies in the ground states and the  entanglement measures for each RAS scheme, we compute the entanglement measures $\mathscr{E}_{L}(\rho)$ and $\mathscr{E}_{VN}(\rho)$ defined in Eqs.~\eqref{eq:el} and~\eqref{eq:evn}; see also Eqs.~\eqref{eq:el_max} and~\eqref{eq:evn_max} for the maximum values of these measures. We plot these measures as a function of the correlation energies. 

\begin{figure}
  \includegraphics{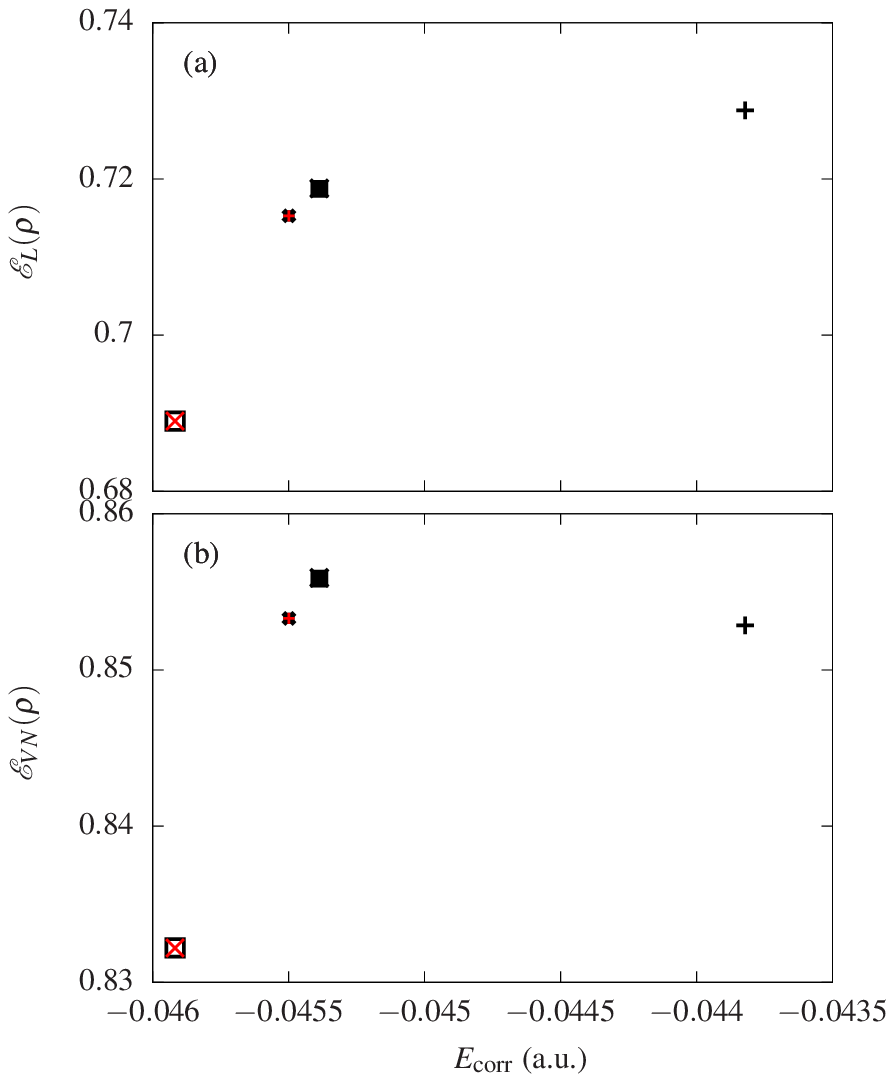}
  \caption{\label{fig:fig2} (a) $\mathscr{E}_L(\rho)$  and (b) $\mathscr{E}_{VN}(\rho)$ and correlation energy, $E_\text{corr}$, for Be with a core. Note that $0\le\mathscr{E}_L(\rho)\le 8\left(1-\frac{4}{M_\text{}} \right) $ and $0\le\mathscr{E}_{VN}(\rho)\le 4\ln\left(\frac{M_\text{}}{2} \right)$ with $M$ the number of spatial orbitals (see Eqs.~\eqref{eq:el_max} and~\eqref{eq:evn_max}). The RAS schemes shown and their corresponding number of configurations are $(M_0,M_1,M_2)(\#\text{configurations})=(1,4,0)(16)[+]$, $(1,5,0)(25)[\times]$, $(1,8,0)(64)[*]$, $(1,13,0)(169)[\boxdot]$,  $(1,1,4)$-D(10)$[\blacksquare]$, $(1,1,7)$-D(50)$[\textcolor{red}{+}]$, $(1,1,12)$-D(145)$[\textcolor{red}{\times}]$. Some points in the plot are superimposed, specifically, $(1,1,4)$-D and $(1,5,0)$; $(1,1,7)$-D and $(1,8,0)$; $(1,13,0)$ and $(1,1,12)$-D.}
  \end{figure}

\begin{figure}
  \includegraphics{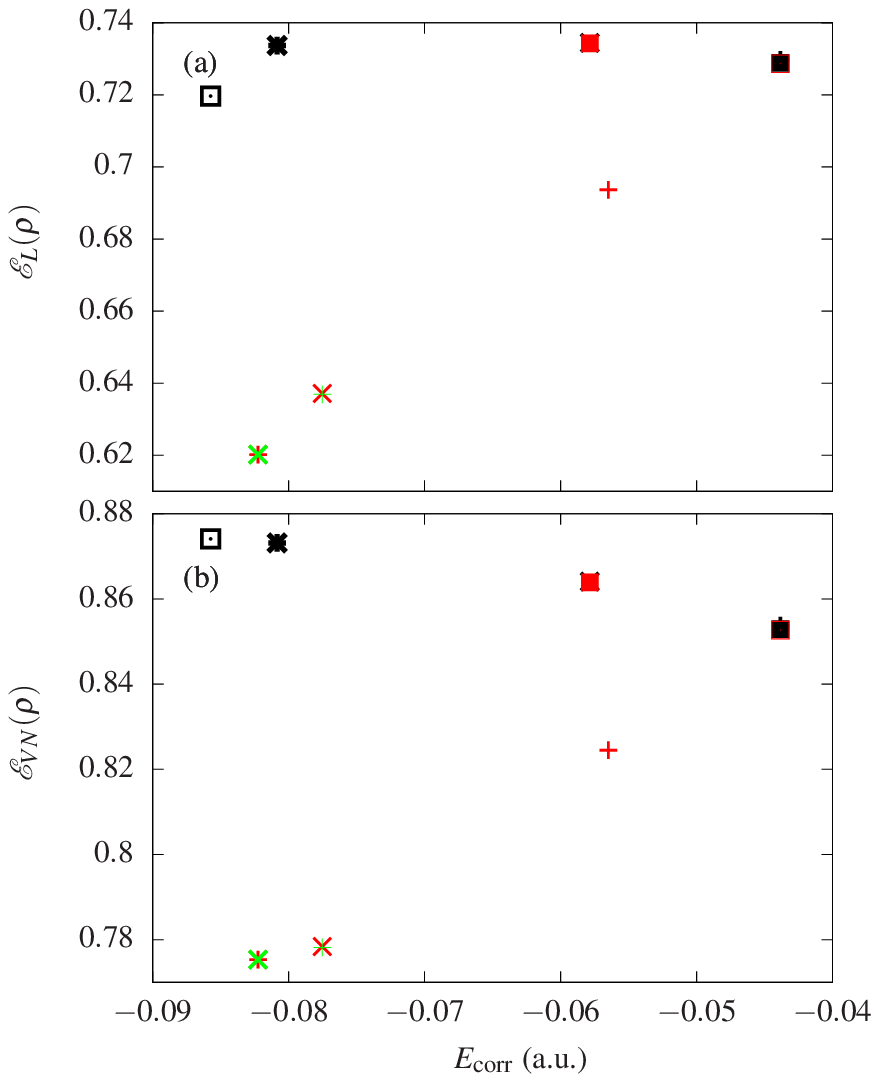}
  \caption{\label{fig:fig3} (a) $\mathscr{E}_L(\rho)$ and (b) $\mathscr{E}_{VN}(\rho)$ and correlation energy, $E_\text{corr}$, for Be without core. Note that $0\le\mathscr{E}_L(\rho)\le 8\left(1-\frac{4}{M_\text{}} \right)$ and $0\le\mathscr{E}_{VN}(\rho)\le 4\ln\left(\frac{M_\text{}}{2} \right)$  with $M$ the number of spatial orbitals (see Eqs.~\eqref{eq:el_max} and~\eqref{eq:evn_max}). The RAS schemes shown  and their corresponding number of configurations are $(M_0,M_1,M_2)(\#\text{configurations})=(0,5,0)(100)[+]$, $(0,6,0)(225)[\times]$, $(0,9,0)(1296)[*]$, $(0,14,0)(8281)[\boxdot]$, $(0,2,3)$-D(43)$[\blacksquare]$, $(0,2,4)$-D(77)$[\textcolor{red}{+}]$, $(0,2,7)$-D(239)$[\textcolor{red}{\times}]$, (0,2,12)-D(709)$[\textcolor{red}{*}]$, $(0,2,3)$-SD(53)$[\textcolor{red}{\boxdot}]$, $(0,2,4)$-SD(93)$[\textcolor{red}{\blacksquare}]$, $(0,2,7)$-SD(267)$[\textcolor{green}{+}]$ and (0,2,12)-SD(757)$[\textcolor{green}{\times}]$. Some points in the plot are superimposed, specifically, $(0,2,3)$-D and $(0,2,3)$-SD; $(0,2,4)$-SD and $(0,6,0)$; $(0,2,7)$-D and $(0,2,7)$-SD; $(0,2,12)$-D and $(0,2,12)$-SD.}
  \end{figure}

First, we analyze the entanglement for Be, and show the results in Figs.~\ref{fig:fig2} and~\ref{fig:fig3}. For both measures, we observe that the wave functions are clearly entangled since the entanglement measures attain values larger than zero~\cite{Plastino2009}. Specifically, for all the cases under study, $\mathscr{E}_{L}(\rho)>0.62$ and $\mathscr{E}_{VN}(\rho)>0.77$. However, the entanglement does not increase with a decreasing energy as we would have expected. For instance, in Fig.~\ref{fig:fig3} we observe that $\mathscr{E}_{L}(\rho)=0.6372$ for $(0,2,7)$-D and $E_\text{corr}=-0.0775218$, whereas it increases up to $0.7188$ and $E_\text{corr}=-0.0453873$ for TD-CASSCF with $(1,5,0)$, as shown in Fig.~\ref{fig:fig2}. In this case, the energy is much lower for $(0,2,7)$-D and the accessible number of configurations is 14 times larger. 

\begin{figure}
  \includegraphics{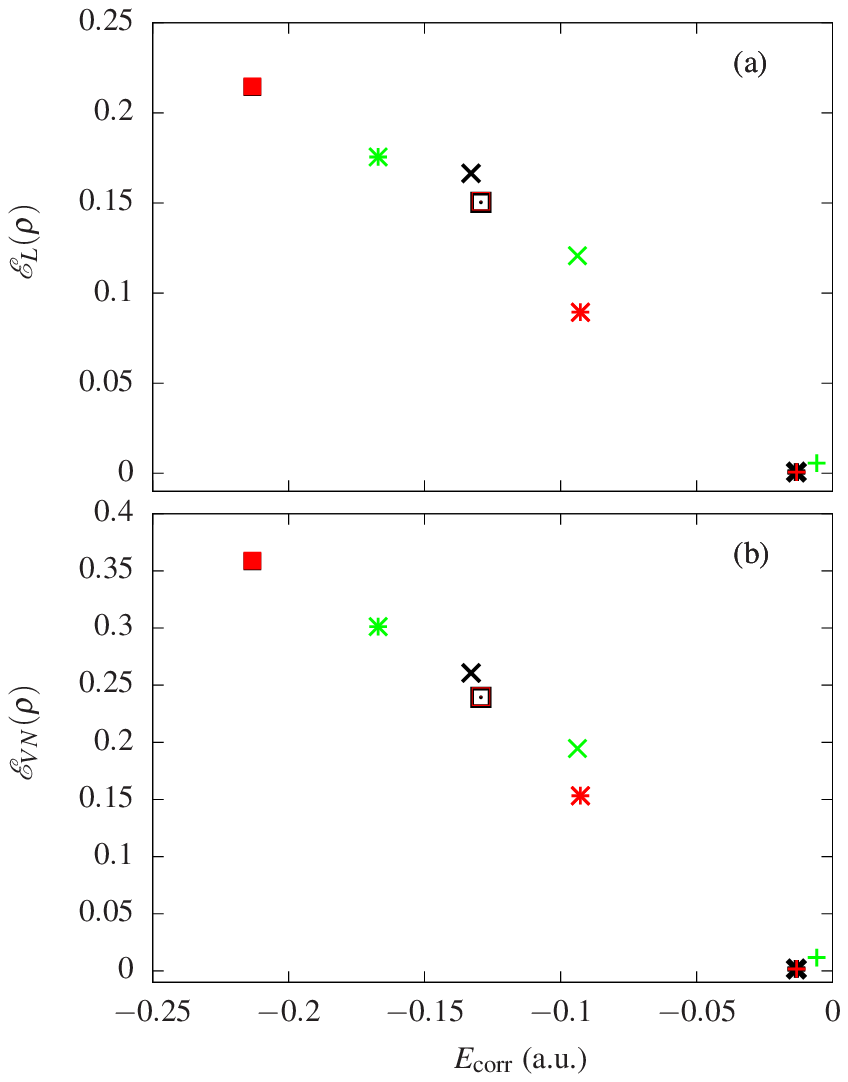}
  \caption{\label{fig:fig4} (a) $\mathscr{E}_L(\rho)$ and (b) $\mathscr{E}_{VN}(\rho)$ and correlation energy, $E_\text{corr}$, for Ne with a core. Note that $0\le\mathscr{E}_L(\rho)\le 20\left(1-\frac{10}{M_\text{}} \right) $ and $0\le\mathscr{E}_{VN}(\rho)\le 10\ln\left(\frac{M_\text{}}{5} \right)$ with $M$ the number of spatial orbitals (see Eqs.~\eqref{eq:el_max} and~\eqref{eq:evn_max}). The RAS schemes shown  and their corresponding number of configurations are $(M_0,M_1,M_2)(\#\text{configurations})=(1,5,0)(25)[+]$, $(1,8,0)(4900)[\times]$, $(1,4,1)$-D(17)$[*]$, $(1,4,4)$-D(329)$[\boxdot]$,  $(1,4,9)$-D(1729)$[\blacksquare]$, $(1,4,1)$-S(9)$[\textcolor{red}{+}]$, $(1,4,4)$-S(33)$[\textcolor{red}{\times}]$, $(1,4,9)$-S(73)$[\textcolor{red}{*}]$, $(1,4,4)$-SD(361)$[\textcolor{red}{\boxdot}]$, $(1,4,9)$-SD(1801)$[\textcolor{red}{\blacksquare}]$, $(2,3,1)$-D(10)$[\textcolor{green}{+}]$, $(2,3,4)$-D(181)$[\textcolor{green}{\times}]$, $(2,3,9)$-D(946)$[\textcolor{green}{*}]$. Note that some points in the plot are superimposed, specifically, $(1,5,0)$, $(1,4,1)$-D and $(1,4,1)$-SD; $(1,4,4)$-D and $(1,4,4)$-SD; $(1,4,4)$-S and $(1,4,9)$-S; $(0,2,12)$-D and $(0,2,12)$-SD.}
  \end{figure}

\begin{figure}
  \includegraphics{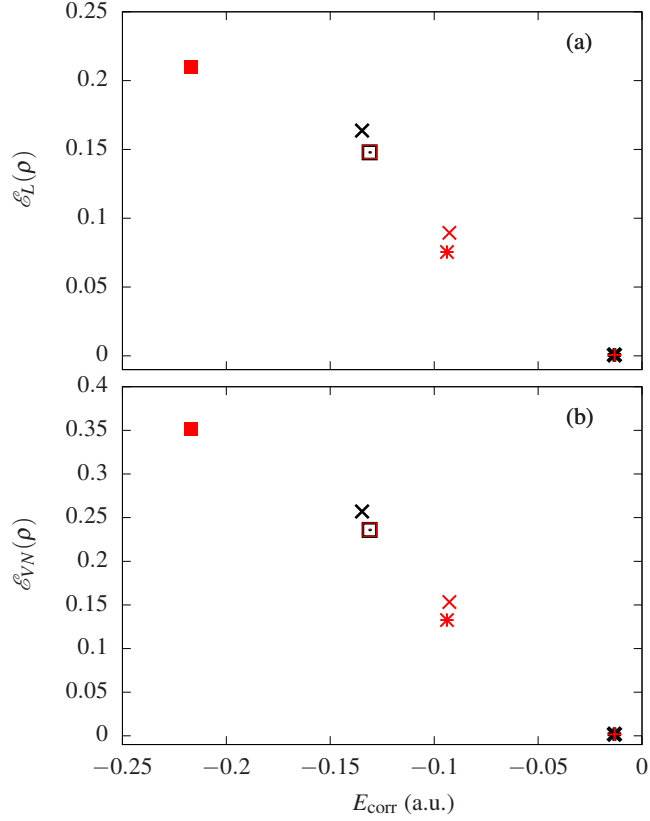}
  \caption{\label{fig:fig5} (a) $\mathscr{E}_L(\rho)$ and (b) $\mathscr{E}_{VN}(\rho)$ and correlation energy, $E_\text{corr}$, for Ne without core. Note that $0\le\mathscr{E}_L(\rho)\le 20\left(1-\frac{10}{M_\text{}} \right)$ and $0\le\mathscr{E}_{VN}(\rho)\le 10\ln\left(\frac{M_\text{}}{5} \right)$ with $M$ the number of spatial orbitals (see Eqs.~\eqref{eq:el_max} and~\eqref{eq:evn_max}). The RAS schemes shown  and their corresponding number of configurations are $(M_0,M_1,M_2)(\#\text{configurations})=(0,6,0)(36)[+]$, $(0,9,0)(15876)[\times]$, $(0,5,1)$-D(26)$[*]$, $(0,5,4)$-D(512)$[\boxdot]$, $(0,5,9)$-D(2746)$[\blacksquare]$, $(0,5,1)$-S(11)$[\textcolor{red}{+}]$ $(0,5,4)$-S(41)$[\textcolor{red}{\times}]$, $(0,5,9)$-S(91)$[\textcolor{red}{*}]$, $(0,5,4)$-SD(36)$[\textcolor{red}{\boxdot}]$, $(0,5,9)$-SD(2836)$[\textcolor{red}{\blacksquare}]$. Some points in the plot are superimposed, specifically, $(0,6,0)$, $(1,5,1)$-S and $(0,5,1)$-D; $(0,5,4)$-D and $(0,5,4)$-SD; $(0,5,4)$-S and $(0,5,9)$-S.}
  \end{figure}

The entanglement measures for Ne are shown in Figs.~\ref{fig:fig4} and~\ref{fig:fig5}. As we have discussed in Sec.~\ref{sec:gs_energy}, the correlation needed to describe the wave function is smaller than for Be. This is indeed reflected in the entanglement measures. For example, for $M_\text{}=6$, $\mathscr{E}_L(\rho)=1.18\times 10^{-3}$ and $\mathscr{E}_{VN}(\rho)=6.88\times 10^{-4}$. Furthermore, we find that for both measures, the entanglement increases as the ground-state energy decreases for fixed $M_0$, attain a maximum for $\mathscr{E}_L(\rho)=0.2097$ and $\mathscr{E}_{VN}(\rho)=0.3519$ for $M_0=0$ and $\mathscr{E}_L(\rho)=0.2147$ and $\mathscr{E}_{VN}(\rho)=0.3590$ for $M_0=1$. We can also observe in Figs.~\ref{fig:fig4} and~\ref{fig:fig5}, that the values of the entanglement measures are slightly larger for wave functions without a core,~\ie, $M_0=0$, than when one orbital is included in the core $\mathcal{P}_0$ subspace, supporting that the measures give reasonable estimates of the degree of entanglement in the system. At the same time it is clear from the numerical values that the measures can not be used to accurately rank different RAS according to their accuracy in obtaining the ground-state energy. For instance, for $(1,4,1)$-D $E_\text{corr}=-0.0132529$ is smaller than $E_\text{corr}=-0.0057859$ for $(2,3,1)$-D, since its contains more configurations. However, its entanglement is smaller, being $\mathscr{E}_{L}(\rho)=6.7628\times 10^{-4}$ compared to $5.69303\times 10^{-3}$ for $(2,3,1)$-D.

We can also use the entanglement measures to evaluate the improvement of TD-RASSCF-D by allowing single excitations in addition to double excitations. In Fig.~\ref{fig:fig3},  we observe that both the ground-state energy and the entanglement for Be for $(0,2,7)$-D and -SD are almost the same, specifically $E=-14.6508198$ and $-14.6508214$, and $\mathscr{E}_L(\rho)=0.6371761$ and $0.6369457$ respectively, showing that the improvement in the energy obtained by including -S is very small. On the other hand, we improve the entanglement of $(0,2,4)$-D by including single excitations and obtain a value that almost coincide with MCTDHF with 6 orbitals. These two situations clearly manifest that for $M_\text{}=6$ the single excitations are important to describe the ground-state wave function, whereas for $M_\text{}=9$ the correlation can be accounted by a larger set of allowed double excitations.

In the case of Ne, we see in Fig.~\ref{fig:fig5} that for $(0,5,1)$-D, by adding single excitations the method is equivalent to MCTDHF, and therefore we obtain identical values for $\mathscr{E}_{L}(\rho)$ and $\mathscr{E}_{VN}(\rho)$ with those two approaches. Even more, TD-RASSCF-D and TD-RASSCF-SD give very similar results for a larger number of orbitals in the $\mathcal{P}_2$ space. This is the case with the RAS schemes $(0,5,4)$, $(0,5,9)$, $(1,4,4)$ and $(1,4,9)$, whose results clearly overlap in Figs.~\ref{fig:fig4} and~\ref{fig:fig5}. Opposed to the case of Be, both entanglement measures increase with decreasing energy, i.e., with the correlation energy. The only exception is the RAS scheme $(2,3,1)$-D, which has a larger energy than MCTDHF with $M=6$, but a larger entropy. Also note that most of the points corresponding to configurations with similar number of orbitals tend to overlap in the plots, therefore, we would expect the same accuracy. However, we observe in Fig.~\ref{fig:fig4} that the points corresponding to $M_0=2$ are isolated, which reflects the lack of relevant configurations compared to the case of $M_0=1$. We will show some further implications of the core space $\mathcal{P}_0$ when describing the photoelectron spectrum in Sec.~\ref{sec:gs_photoelectron}.

 In conclusion, while the relation between the entanglement measures and the correlation energies is as expected for Ne, no monotonic acceptable relation was found for Be. The entanglement measures can therefore not be used in general to address the degree of correlation in a system, and therefore, we do not find it relevant to consider these measures in the next section on  ultrafast photoionization dynamics. 

\subsection{Photoionization dynamics}
\label{sec:gs_photoelectron}

In this section we investigate the effect of correlation on the photoelectron spectrum (PES) after the interaction of Be or Ne with an ultrashort linearly polarized laser pulse. The PES is computed by projecting the propagated time-dependent many-electron wave function on Coulomb scattering functions in the outer region after the end of the laser pulse~\cite{Omiste2017_be}. To induce ionization, we consider pulses given by the vector potential 
\begin{equation}
\label{eq:a_laser}
  \vec{A}(t)=A_0\hat{z}\cos^2[\omega t/(2n_p)]\sin\omega t,
\end{equation}
 where $n_p$ is the number of cycles, $\omega$ is the central photon energy and the duration of the pulse is $T=2\pi n_p/\omega$. The pulse begins at $t=-T/2$. 

First, we focus on the following processes involved in the photoionization of the ground-state of Be~$(1s^22s^2,\,\singlets)$
\begin{eqnarray}
\nonumber
  \text{Be}(1s^22s^2&&,\,\singlets)+\gamma\\
\label{eq:be_1s2s}
&& \rightarrow  \text{Be}^+\left(1s^22s,\,\doubletshalf\right)+e^-(p)\\
\label{eq:be_1s2p}
&& \rightarrow  \text{Be}^+\left(1s^22p,\,\doubletpodd\right)+e^-(s,d)\\
\label{eq:be_1s3s}
&& \rightarrow  \text{Be}^+\left(1s^23s,\,\doubletshalf\right)+e^-(p),
\end{eqnarray}
with threshold energies of 9.32, 13.28 and 20.26~eV, respectively\cite{NIST_ASD} and $\gamma$ in  the first line denoting the photon.
\begin{figure}
  \includegraphics{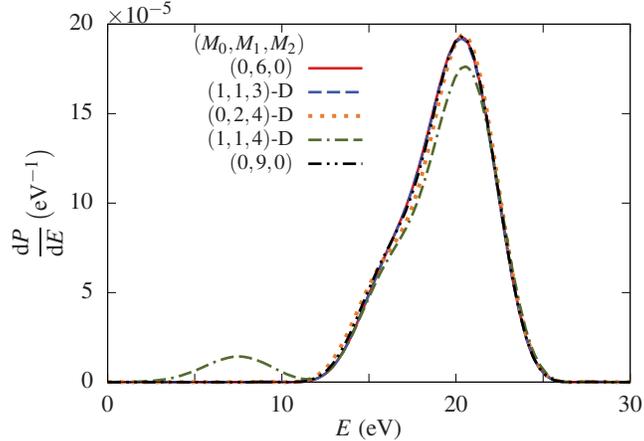}
  \caption{\label{fig:fig6} Photoelectron spectra of Be as a function of the emitted electron energy for a 10 cycle linearly polarized pulse with peak intensity $10^{13}$~W/cm$^2$ for $\omega=30$~eV. Note that results of $(0,2,4)$-SD and MCTDHF with 6 orbitals, i.e., $(0,6,0)$, overlap and only results of the latter approach is shown.}  
\end{figure}
For the different channels, the allowed angular momentum of the ejected electron is determined by the photoionization selection rules, which require that the total term in the final states is an odd $^1\text{P}$ term in LS-coupling.~\cite{Omiste2017_be, Omiste2018_neon}. We briefly describe the dynamics for each case: in the process~\eqref{eq:be_1s2s}, the laser field removes an electron from the $2s$ shell. On the other hand, the dynamics is much more complicated in processes~\eqref{eq:be_1s2p} and~\eqref{eq:be_1s3s}, where one electron is ejected from the $2s$ shell and its counterpart is promoted to an excited shell, $2p$ or $3s$, respectively. As we discussed in Ref.~[\onlinecite{Omiste2017_be}], the correlation in the wave function plays a major role in the description of these latter  processes, for example, they cannot be described by TDHF. In Fig.~\ref{fig:fig6} we show the PES extracted at $t\approx 57$~a.u. after interacting with an ultrashort pulse with a peak intensity $10^{13}$~W/cm$^2$, $n_p=10$ and $\omega=30$~eV for $M\le 6$ including MCTDHF with 9 orbitals to check the accuracy and convergence. Note that we do not show $(0,2,4)$-SD because the results of this RAS scheme overlap with the results of MCTDHF with 6 orbitals. The expected peaks in the PES are located at $20.68$, $16.72$ and $9.74$~eV for the channels~\eqref{eq:be_1s2s}-\eqref{eq:be_1s3s}, respectively. As in our earlier study~\cite{Omiste2017_be}, we observe in Fig.~\ref{fig:fig6} one peak at approximately $20$~eV in the PES when using TD-RASSCF-D without core and MCTDHF, corresponding to the process~\eqref{eq:be_1s2s}. We also find a shoulder around 17~eV corresponding to an electron ejected in process~\eqref{eq:be_1s2p}. This channel can, however, not be clearly resolved due to the large bandwidth of the short pulse. The position of the maximum of the PES is in good agreement for the RAS schemes $(0,6,0)$ and $(0,2,4)$-D, but they slightly differ in the shoulder, due to a shift in the energy of the ejected $s/d$ electron coming from the $\text{Be}^+\left(1s^22p,\,\doubletpodd\right)$ channel.
\begin{figure}
  \includegraphics{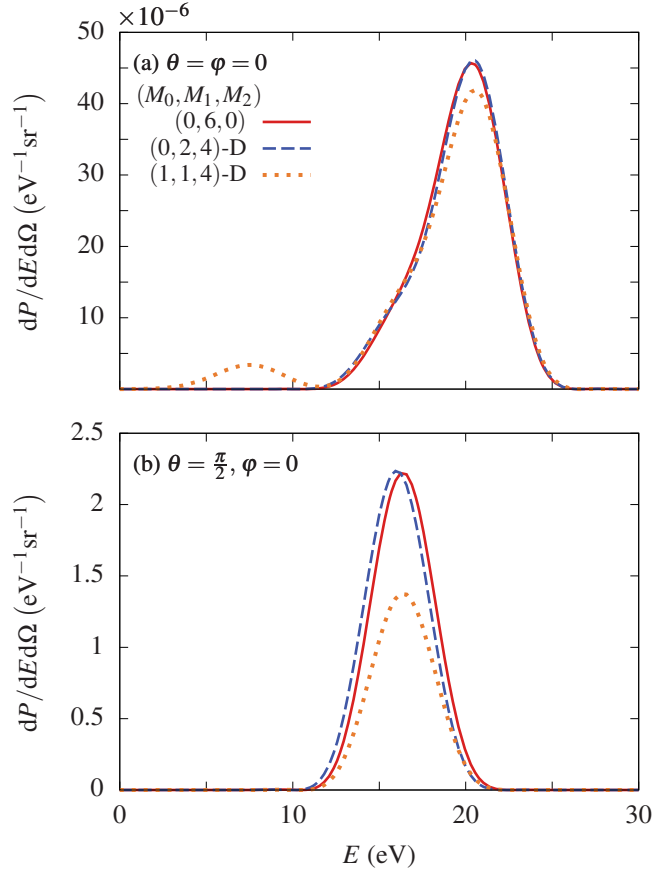}
  \caption{\label{fig:fig7} Triply differential probabilities of Be (a) parallel $(\theta=0)$ and (b) perpendicular $(\theta=\pi/2,\varphi=0)$ to the polarization axis of the laser pulse
 as a function of the emitted electron energy for a 10 cycle linearly polarized pulse with peak intensity $10^{13}$~W/cm$^2$ for $\omega=30$~eV.  Note that results of $(0,2,4)$-SD and MCTDHF with 6 orbitals, i.e., $(0,6,0)$, overlap and only results of the latter approach is shown.}  
\end{figure}

Furthermore, even if we restrict the correlation by adding an orbital to the core, as in $(1,1,3)$-D, we observe that the PES is almost unaltered. This is, however, not a general trend for the system, since for $M_0=1$ and $M_\text{}\ge 6$ we find an extra peak at approximately $7.5$~eV. For example, for $(1,1,4)$-D, this new peak is located at approximately 7.52~eV and it is associated with an ionization threshold of 22.48~eV, as we see in Fig.~\ref{fig:fig6}.  We can identify the channel of the Be$^+$ ion corresponding to these peaks by analyzing the triply differential probabilities (TDP) of the photoelectrons parallel and perpendicular to the polarization axis of the laser. The results of the TDPs are shown in Fig.~\ref{fig:fig7}. We consider three examples with the same number of orbitals to illustrate the role of the different RAS. We find that this peak at $7.5$~eV is not observed perpendicular to the polarization axis of laser [Fig.~\ref{fig:fig7}(b)] but only parallel to the polarization [Fig.~\ref{fig:fig7}(a)]. This angular pattern is consistent with the ejection of a $p$ electron with $m=0$. This indicates that the ionization occurs through the $\text{Be}^+\left(1s^23s,\,\doubletshalf\right)$ channel [process~\eqref{eq:be_1s3s}], but the peak is shifted by approximately 2.22~eV with respect to the exact value~\cite{NIST_ASD}. The abscence of the peak at $7.5$~eV for $M\le 5$ indicates that the system is not able to reproduce accurately the $\text{Be}^+\left(1s^23s,\,\doubletshalf\right)$ ionization channel, whereas for $M\ge 6$ it is possible, since the sixth orbital is the $3s$ of the ground-state wave function by construction. However, the restriction imposed by the core does not allow to account accurately for its role in  the wave function and the subsequent dynamics, overestimating the ionization through this channel. Furthermore, the main peak is shifted to higher energies,~\ie, the electron is less bound, and the ionization probability is lower. By setting $M_0=0$, as in the case of $(0,2,4)$-D, we recover the same behavior observed for MCTDHF with 6 and 9 orbitals.  Furthermore, we also observe in Fig.~\ref{fig:fig7}(b) that the ionization through  $\text{Be}^+\left(1s^22p,\,\doubletpodd\right)$ is underestimated for $(1,1,4)$-D compared with $(0,2,4)$-D and MCTDHF with 6 orbitals. Let us remark that the results of the computation for TD-CASSCF $(1,5,0)$ are indistinguishable from the $(1,1,4)$ -D results.

Next, we consider the photoionization of the ground-state of Ne~$(1s^22s^22p^6,\,\singlets)$,
\begin{eqnarray}
\nonumber
  \text{Ne}(1s^22s^22p^6&&,\,\singlets)+\gamma \\
\label{eq:ne_2p}
&& \rightarrow  \text{Ne}^+\left(1s^22s^22p^5,\,\doubletpodd\right)+e^-(s,d)\\
\label{eq:ne_2s}
&& \rightarrow  \text{Ne}^+\left(1s^22s2p^6,\,\doublets\right)+e^-(p)
\end{eqnarray}
where the ionization thresholds are 21.56 and 48.48~eV for the processes~\eqref{eq:ne_2p} and \eqref{eq:ne_2s}, respectively~\cite{NIST_ASD}.

\begin{figure}
  \includegraphics{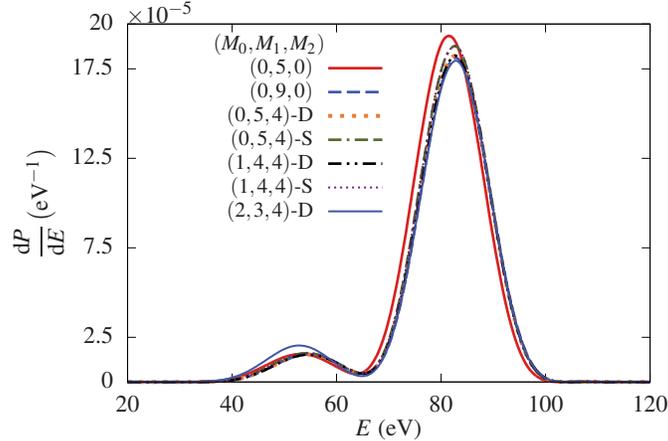}
  \caption{\label{fig:fig8}  Photoelectron spectra of Ne as a function of the emitted electron energy for a 10 cycle linearly polarized pulse with peak intensity $10^{14}$~W/cm$^2$ for $\omega=105$~eV. Note that results of $(0,5,1)\text{-D},(0,5,1)\text{-S},(0,5,1)\text{-SD},(1,4,1)\text{-S},(1,4,1)\text{-D}$ and $(0,6,0)$ overlap with $(0,5,0)$ and $(1,8,0)$; $(0,5,4)$-SD with $(0,9,0)$.}  
\end{figure}

 To understand the role of single and double excitations, we show in Fig.~\ref{fig:fig8} the PES for $\omega=105$~eV and 10 cycles for different levels of approximation of TD-RASSCF. The peaks are expected at $56.52$ and $83.44$~eV, and this expectation is reproduced with good accuracy by all RAS schemes shown~\cite{Omiste2018_neon}. Let us remark that the PES corresponding to $(M_0,M_1,M_2)=(0,5,0),(0,5,1)\text{-D},(0,5,1)\text{-S},(0,5,1)\text{-SD},(1,4,1)\text{-S}$, $(1,4,1)\text{-D}$ and $(0,6,0)$ are indistinguishable in Fig.~\ref{fig:fig8}, as well as the ones corresponding to $(1,8,0)$ and $(0,5,4)$-SD with $(0,9,0)$. Therefore, we only plot one of each group.  We do show -SD because the only relevant case for $M\le 9$,~\ie, $(0,5,4)$-SD overlaps $(0,5,4)$-D. The fact that the PES for $(1,8,0)$ and $(0,9,0)$ are almost identical reveals that due to the high nuclear charge, the inner electrons are tightly bound, hence, their excitations are very unlikely to play a role for the photon energies considered here. The implications of the presence of a less bound core electrons for the ground state were discussed in Sec.~\ref{sec:gs_energy}. As we see in Fig.~\ref{fig:fig8}, the main peak is shifted to higher energies for $M=9$, with the maximum located at the same energy for $(0,5,4)$-D, $(1,4,4)$-D, $(0,5,4)$-S and $(1,4,4)$-S. However, we find that the signal is larger for $(1,4,4)$-S than for $(0,5,4)$-D, which is again a bit higher than $(1,8,0)$ and MCTDHF with 9 orbitals. This means that the energy levels of Ne$^+$ are well described, independently of the number of excitations allowed, but the dynamics, which is responsible of the weight of each channel in the photoionization, is sensitive to the RAS used. On the other hand, if we set $M_0=2$, the peak corresponding to the removal of an electron from the $2s$,~\ie,  $\text{Ne}(1s^22s^22p^6,\,\singlets)\rightarrow\text{Ne}^+\left(1s^22s2p^6,\,\doublets\right)+e^-(p)$, is overestimated and shifted to lower energies, but the high energy peak is still in agreement with the MCTDHF with 9 orbitals. Let us remark that, opposed to the case of Be, we do not observe extra peaks in the PES, because the other ionic channels of Ne$^+$ associated to excitations from $2s$ are above the central photon energy considered, in this case $105$~eV. These findings show that the excitation from the $2s$ subshell is important to reproduce the photoinduced dynamics in Ne and, as we have discussed in Sec.~\ref{sec:gs_energy}, the nuclear charge is not strong enough to tightly bind $2s$ electrons. 

\section{Summary and Conclusions}
\label{sec:conclusions}

We have explored the effects of electron-electron correlation for the ground states in Be and Ne as well as for ultrafast photoionization dynamics induced by attosecond pulses. We have used the TD-RASSCF methodology with different active orbital spaces and excitation level - singles, doubles and singles and doubles. In particular, we have investigated the implications of including an always occupied dynamical core space $\mathcal{P}_0$ in the RAS. First, we found that the effects on the ground-state energy and correlation of Be are more significant than for Ne. For instance, for Be, we found that the inclusion of one spatial orbital in the core strongly affects the correlation energy, which seems to converge to $0.04$ above the exact correlation energy. In the case of Ne, we only observed the same limitation when including two spatial orbitals in the core. This difference is because the nuclear charge of Ne is larger, implying that the inner electrons are more tightly bound than in Be and the many-body wavefunction can be accurately reproduced using a core with one spatial orbital. Second, the dynamical core also affects the ultrafast photoionization dynamics. For example, in the case of Ne, the peak in the PES corresponding to removing one electron from the $2s$ subshell was found to be exaggerated, because of the explicit restriction of the excitation from this orbital in the configurational space. Therefore, the photoionization process can only be driven by the dynamics of the $2s$ orbital. Furthermore, in the case of Be, we observed that the photoionization accompanied by the excitation of one electron from the $2s$ subshell to the $3s$ has a nonvanishing probability when fixing the $1s$ orbital in the core space, whereas this probability was negligible without a core. A similar effect was observed in Refs.~\onlinecite{Pabst2012,Chen2015} by switching on and off excitations from particular subshells in the TD-CIS of larger atoms. 

In connection with the ground-state studies, we compared the values for the correlations energies with the values of the linear and von Neuman entropy entanglement measures for several RAS schemes. We found that the entropies do not always increase with the magnitude of the correlation energy and the number of orbitals or configurations, showing that they are in general not good entanglement measures for many-body systems, and should only be considered as separability criteria. Overall, the measures $\mathscr{E}_L(\rho)$ and $\mathscr{E}_{VN}(\rho)$ work much better for Ne than for Be. The origin of this difference is not clear at present.

This work together with the previous investigations on beryllium~\cite{Omiste2017_be} and neon~\cite{Omiste2018_neon} manifest the strength and flexibility of the TD-RASSCF method to describe ultrafast photoionization in atoms. Future implementation of surface flux and related methods~\cite{Scrinzi2010,Scrinzi2012,Morales2016} will allow efficient computation of  photoelectron spectra and photoelectron momentum distributions for non linearly-polarized laser sources in many-electron atoms, which so-far have only been predicted in  helium~\cite{NgokoDjiokap2014,NgokoDjiokap2015,NgokoDjiokap2016,Djiokap2017}.

\begin{acknowledgments}
 This work was supported by the Villum Kann Rasmussen (VKR) Center of Excellence QUSCOPE and by NSERC Canada (via a grant to Prof. P. Brumer). The numerical results presented in this work were obtained at the Centre for Scientific Computing, Aarhus.
\end{acknowledgments}

\appendix

\section{Conservation of magnetic quantum number in TD-RASSCF for arbitrary number of excitations}
\label{sec:conservation_m}
We prove that the magnetic quantum numbers of each orbital are conserved by the EOM of TD-RASSCF for any number of excitations, when assuming that the total quantum magnetic quantum number of each configuration, $M_L$, and each orbital, $m_i$, are well defined. In Ref.~[\onlinecite{Omiste2018_neon}] we showed that this is the case for MCTDHF and TD-RASSCF with even number of excitations. These cases differ from TD-RASSCF with arbitrary number of excitations in the term $\dot\rho_{i'}^{j''}(t)$ on the right hand side of Eq.~\eqref{eq:p_space}. Then, to extend the proof to TD-RASSCF with an arbitrary number of excitations we first rewrite the $\mathcal{P}$-equation as
\begin{widetext}
  \begin{equation}
    \label{eq:p_space_eta}
    \eta_{l'}^{k''}(t)=\sum_{j''i'}\left\{-\dot\rho_{i'}^{j''}(t)-i\sum_{klm}\left[v_{kl}^{j''m}(t)\rho_{i'm}^{kl}(t)-v_{i'm}^{kl}(t)\rho_{kl}^{j''m}(t)\right]\right\}\left[A^{-1}\right]_{j''l'}^{i'k''}(t)-ih_{l'}^{k''}(t),
  \end{equation}
\end{widetext}
where $\left[A^{-1}\right]_{j''l'}^{i'k''}(t)$ is the inverse of $A_{k''i'}^{l'j''}(t)$ taking into account that $A$ maps the space $(i'j''$ to $(l'k'')$~\cite{Madsen2018}. We will also use that the time derivative of the one-body density is
\begin{eqnarray}
\nonumber
  \dot\rho_{i'}^{j''}(t)&=&\langle\Psi(t)|c_{i'}^\dagger c_{j''}|\Psi(t)\rangle=\\
 \label{eq:rho_expand} 
  &=& \sum\limits_{\mathbf{I}\in\mathcal{V}_{RAS}}\left[\dot C_\mathbf{I}(t)^\star\escalar{\Phi_\mathbf{I}}{\Psi_{j''}^{i'}}+\escalar{\Psi_{i'}^{j''}}{\Phi_\mathbf{I}}\dot C_\mathbf{I}(t)\right],
\end{eqnarray}
where the one-particle-one-hole states $\bra{\Psi_{i'}^{j''}}=\bra{\Psi}c_{i'}^\dagger c_{j''}$ and  $\ket{\Psi_{j''}^{i'}}=c_{i'}^\dagger c_{j''}\ket{\Psi}$ are used. 
Let us first assume that $m_{i'}\ne m_{j''}$. On the one hand, $\ket{\Psi}$ and the configurations $\ket{\Phi_\mathbf{I}}$ with non-zero amplitude have a magnetic quantum number $M_L$ by construction. On the other hand, $\ket{\Psi_{j''}^{i'}}$ and $\bra{\Psi_{i'}^{j''}}$ have a magnetic quantum number $M_L-m_j+m_i$ and $M_L-m_i+m_j$, respectively. Therefore, $\ket{\Phi_\mathbf{I}}$, $\ket{\Psi_{i'}^{j''}}$ and $\bra{\Psi_{j''}^{i'}}$ have different magnetic quantum numbers, thus, $\escalar{\Phi_\mathbf{I}}{\Psi_{j''}^{i'}}=\escalar{\Psi_{i'}^{j''}}{\Phi_\mathbf{I}}=0$ and therefore $\dot\rho_{i'}^{j''}(t)= 0$. Then,~\eqref{eq:p_space_eta} is the same than in the case of even excitations, implying that terms with $m_{i'}\ne m_{j''}$ do not mix the magnetic quantum numbers in $\eta_{i'}^{j''}$~[\onlinecite{Omiste2018_neon}]. 

To complete the proof we now consider $m_{i'}=m_{j''}$. As in the previous case, Eqs.~\eqref{eq:ci_dot} and~\eqref{eq:q_space} do not modify $m_i$'s or $M_L$~[\onlinecite{Omiste2018_neon}]. Then, we evaluate the term  $\sum_{j''i'}\dot\rho_{i'}^{j''}(t)\left[A^{-1}\right]_{j''l'}^{i'k''}(t)$ in Eq.~\eqref{eq:p_space_eta}, since it is the only contribution which differs from the case of even excitations. $A_{k''i'}^{l'j''}(t)= 0$ if $m_{j''}\ne m_{k''}$ or $m_{i'}\ne m_{l'}$ as it is shown in Ref.~[\onlinecite{Omiste2018_neon}], therefore $\left[A^{-1}\right]_{j''l'}^{i'k''}(t)= 0$ if $m_{j''}\ne m_{k''}$ or $m_{i'}\ne m_{l'}$. In addition, since $\dot\rho_{i'}^{j''}(t)$ may not be zero only for $m_{i'}=m_{j''}$ we can conclude that $\sum_{j''i'}\dot\rho_{i'}^{j''}(t)\left[A^{-1}\right]_{j''l'}^{i'k''}(t)=0$,~\ie, $\eta_{l'}^{k''}=0$, if $m_{l'}\ne m_{k''}$ implying that the $\mathcal{P}$-space equation~\eqref{eq:p_space} does not mix $M_L$ or $m_i$'s. 


%

\end{document}